# Atomic arrangement of van der Waals heterostructures using X-ray scattering and crystal truncation rod analysis

Ryung Kim[a†], Byoung Ki Choi[b,c†], Kyeong Jun Lee[a], Hyuk Jin Kim[b], Hyun Hwi Lee[d], Tae Gyu Rhee[b,e], Yeong Gwang Khim[b,e], Young Jun Chang[b,e], and Seo Hyoung Chang[a*]

[a]*Department of Physics, Chung-Ang University, Seoul 60974, Republic of Korea*
[b]*Department of Physics, University of Seoul, Seoul 02504, Republic of Korea*
[c]*Advanced Light Source (ALS), E. O. Lawrence Berkeley National Laboratory, Berkeley, California 94720, USA*
[d]*Pohang Accelerator Laboratory (PAL), Pohang University of Science and Technology (POSTECH), Pohang 37673, Republic of Korea*
[e]*Department of Smart Cities, University of Seoul, Seoul 02504, Republic of Korea*

**Vanadium diselenide ($VSe_2$) has intriguing physical properties such as unexpected ferromagnetism at the two-dimensional limit. However, the experimental results for room temperature ferromagnetism are still controversial and depend on the detailed crystal structure and stoichiometry. Here we introduce crystal truncation rod (CTR) analysis to investigate the atomic arrangement of bilayer $VSe_2$ and bilayer graphene (BLG) heterostructures grown on a 6H-SiC(0001) substrate. Using non-destructive CTR analysis, we were able to obtain electron density profiles and detailed crystal structure of the $VSe_2$/BLG heterostructures. Specifically, the out-of-plane lattice parameters of each $VSe_2$ layer were modulated by the interface compared to that of the bulk $VSe_2$ 1T phase. The atomic arrangement of the $VSe_2$/BLG heterostructure provides deeper understanding and insight for elucidating the magnetic properties of the van der Waals heterostructure.**

e-mail: cshyoung@cau.ac.kr

# 1. Introduction

Transition metal dichalcogenides (TMDCs) have been intensively investigated due to their intriguing physical properties at the two-dimensional (2D) limit [1–3]. Because of their weak van der Waals (vdW) interactions compared to covalent bonds, TMDCs and graphene are known as ideal platforms for investigating intriguing 2D physics and fabricating heterostructures, even when controlling twisted angles [4,5]. For instance, Mott insulating state and superconductivity of twisted bilayer graphene emerged from strongly correlated electrons [6]. Moreover, the charge-density wave (CDW) and magnetic properties of the heterostructures can be modulated by interlayer coupling, reduced dimensionality, and thermal fluctuations [7]. Understanding the interfacial structure of TMDC-based heterostructures is important for exploring novel phenomena and engineering their functionalities.

Compared to the paramagnetic state of the Vanadium diselenide ($VSe_2$), unexpected ferromagnetism at room temperature of monolayer 1T $VSe_2$ has been experimentally observed and extensively investigated. However, scanning tunneling microscopy (STM) and angle-resolved photoemission spectroscopy (ARPES) have measured CDW phases with $4 \times 4 \times 3$ periodicity of bulk and distinct periodicity of 2D, which can normally prohibit the appearance of ferromagnetic orders [8,9]. In addition, there have been inconsistencies in numerous experiments on the magnetic moments, which are believed to be due to the quality of samples, e.g., different cation / anion stoichiometries, strain, and crystal structures, and have been controversial to date [10,11].

The interfacial effects of $VSe_2$ heterostructures can modify electronic structures and are closely related to exotic physical properties such as enhanced magnetic moments and flat bands [11,12]. However, the understanding of the interfacial structure of $VSe_2$ heterostructures is still not satisfactory due to experimental technical difficulties. Therefore, resolving atomic-level interfacial structures in $VSe_2$-based heterostructures is very critical for unveiling and determining their intrinsic physical properties.

Here, we investigate the atomic arrangement of bilayer VSe$_2$/bilayer graphene (BLG) heterostructure using surface X-ray scattering and crystal truncation rod (CTR) analysis. To determine the out-of-plane and in-plane lattice parameters of the bilayer VSe$_2$/BLG heterostructure, a model system was constructed through simulation and fitting analysis based on the Fourier transformation of complex structure factors. We then obtained appropriate electron density profiles of bilayer VSe$_2$, BLG, and reconstructed SiC substrates from the samples, which is consistent with the previous study [13]. Resonant energy dependence and off-specular CTR analysis were used to confirm detailed in-plane lattice information of the bilayer VSe$_2$/BLG heterostructure.

## 2. Experiment

The high-quality epitaxial thin films were grown on 6H-SiC (001) substrates (Crystal Bank at Pusan National University) and fabricated using a home-built molecular beam epitaxy (MBE) deposition. The substrate was annealed to form BLG on the Si. VSe$_2$ ultrathin films were grown on BLG with a base pressure of $2 \times 10^{-10}$ Torr. V (99.8%) and Se (99.999%) were simultaneously evaporated by an electron beam evaporator (EFM3, Omicron GmBH) and a Knudsen cell (Effucell Co.). A deposition temperature of 250°C was used and film growth was monitored by reflection-high energy electron diffraction (RHEED). The growth rate was 20 min per monolayer and the samples were post-annealed at 350°C for 30 min. Detailed growth conditions were described elsewhere [14].

X-ray diffraction (XRD) measurements were performed using a laboratory X-ray source with Cu K$_\alpha$ radiation and a high-resolution X-ray diffractometer (Bruker AXS D8). For high resolution, Göbel mirror and Ge(004) two bounce monochromatic setup were used on the incident beam side. X-ray scattering studies were performed using a synchrotron radiation source at beamline 5A of Pohang Light Source II (PLS II, South Korea). The typical size of the focused beam was about 700×50 μm$^2$. A six-circle Huber goniometer with a 2D detector (PILATUS II 200K, Dectris) was used for the higher reciprocal lattice spaces and resolution. The X-ray energies for specular and off-specular rod scans were 11.85 keV and 12.55 keV, respectively, as shown in Fig. 2 and Fig. 4, respectively. A Matlab program and a python-based program (e.g., jupyter notebook) were used for 2D detector data analysis and the fitting calculations, respectively. Detailed fitting calculation was described in the Section 1 in the Supplementary Information.

## 3. Results and discussion.

Figure 1(a) shows a schematic of a 1T VSe$_2$/BLG heterostructure. VSe$_2$ mainly exhibits two polymorphs: 1T phase and 2H phase. The 1T structure refers to an octahedral in which transition metal (V) is sandwiched between the chalcogen layer (Se) and the reversed chalcogen layer, and the 2H phase has two symmetrical equilateral triangles of chalcogen layers. Although theoretical studies proposed two polymorphs for room temperature ferromagnetism, the 1T VSe$_2$ phase has been reported in most experiments [15–17]. Using X-ray diffraction and TEM measurements, single crystals and thin films of several nanometers are known to be 1T VSe$_2$ phases. In particular, our monolayer and bilayer VSe$_2$ samples fabricated with MBE were identified as 1T VSe$_2$ phases using STM [14].

Atomic-level crystal structure verification of VSe$_2$ heterostructure samples is a prerequisite for understanding physical properties such as unsolved ferromagnetic orders, but it is still difficult using non-destructive experimental methods. Bulk crystal of 1T VSe$_2$ is shown in Fig. 1(b) with the lattice parameter with a=b=3.3 Å, c=6.1 Å, γ=120° (space group P$\bar{3}$m1) [17,18]. Using the TEM, the out-of-plane lattice parameter of VSe$_2$ nanosheets can be varied up to 5.88 Å, which was significantly different from that of the bulk samples. STM was only able to confirm in-plane lattice parameters, 3.2–3.4 Å [14,19]. To identify the lattice parameters of the bilayer VSe$_2$ heterostructure sample, we attempted to perform XRD measurements using a laboratory X-ray source as shown in Fig. 1(c). Our VSe$_2$ heterostructure sample consists of a bilayer VSe$_2$ (approximately 1.2 nm), bilayer graphene (~0.6 nm), and a 500 μm thick SiC substrate. Figure 1(c) demonstrates that the scattering pattern of bilayer VSe$_2$ heterostructure samples was exactly same as that of the SiC substrate because of the weak scattering cross-section in the XRD measurement.

We investigated the buried interfacial structures of the VSe$_2$ heterostructure as shown in Fig. 2 by performing synchrotron-based X-ray scattering with a beam energy of 11.85 keV. Conventional

XRD usually measures the average value of the lattice parameters of a thin film. Here, we introduce CTR scattering, a powerful tool for resolving the interfacial structure and lattice parameters of individual layers at the sub-Å scale [20–22]. CTR is a diffraction patterns derived from the crystalline surface of a high-quality sample, indicating that there is a scattering intensity between typical Bragg peaks detectable by conventional XRD due to the Fourier transformation with truncations or broken periodicity [23]. To clearly detect these weak signals in the collected intensities, background subtraction and careful geometrical correction were required. Then, by fitting the CTR scattering with or without a model, detailed interfacial structures can be determined with atomic-level resolution. Note that CTR scattering can also use off-specular rod scans for resolving in-plane lattice information. Here, we performed a specular rod scan along the (00L) direction in the reciprocal lattice space to confirm the surface and interfacial structures of $VSe_2$ heterostructures.

To describe the interfacial structure with atomic-level resolution, we measured the CTR scattering of monolayer and bilayer $VSe_2$ samples, as shown in Fig. 2(a). Our monolayer and bilayer samples have some domains with different thicknesses, but they are named as major domains of nominal thickness [see Fig. S1 and Fig. S2 in the Supplementary Information]. The CTR data of the monolayer $VSe_2$ heterostructure was distinguishable by comparison with the reported CTR result of the bilayer graphene on SiC [13]. The main features due to the monolayer $VSe_2$ heterostructure were dip features at around L = 7.733 and subsequential peaks, indicated by (c) in Fig. 2(a). Evidently, the diffraction pattern of the substrate (half-infinite periodic layers) produces features similar to delta functions, especially in the vicinity of SiC(0006) and SiC(00012).

To explore the thickness dependence of $VSe_2$, we compared the CTR data of bilayer $VSe_2$ samples, indicating that the intensity of $VSe_2(002)$ is clearly stronger than that of monolayer $VSe_2$ samples. Conventional XRD analysis obtained from $VSe_2(002)$ reflection in bilayer $VSe_2$ provided that the out-of-plane lattice parameter was approximately 6.13 ± 0.1 Å. However, the value was not layer-resolved but the average of lattice parameters obtained from monolayer and bilayer $VSe_2$. In particular, the CTR data represents broad peaks with relatively weak intensities originated from the interference between $VSe_2$, graphene, and the surface modulation of SiC substrate. Therefore, to obtain an atomic-level accurate interfacial structure, CTR analysis using a fine-fitting model is required. Then, we can resolve the lattice parameters of each VSe$_2$ layer.

For the scattering vector along the 00L direction, the diffracted CTR signal and Bragg peak can be detected using a pixel 2D area detector. Since CTR intensity occur when the surface is truncated, the CTR signal is highly dependent on the dimensionality and modulation of surface structure. As shown in Fig. 2(b), area detector near the Bragg reflection, denoted by (b) in Fig. 2(a), can identify the CTR intensity with typical Gaussian shape (red box) and Bragg-related peak (white box) along the 00L direction in the reciprocal lattice space [24]. The CTR signal was maintained up to (0014) and showed the anti-Bragg reflection, denoted by (c) in Fig. 2(a) and Fig. 2(c), respectively. This indicates that the VSe$_2$ heterostructure was a high-quality sample with a well-defined SiC surface and the following interface. Figure 2(d) shows stronger CTR intensity compared to monolayer VSe$_2$, indicating a clear thickness dependence of the CTR signal of the VSe$_2$ heterostructure.

To gain further insight into the detailed atomic arrangement of bilayer VSe$_2$/BLG heterostructure, simulation and CTR fitting were performed based on the scattering factor, $F_{total}$, as shown in Fig. 3(a). The CTR intensity is proportional to $F_{total}$ and can be calculated as following

$$F_{total} = \sum_j \theta_j \left(f_j + f' + i \cdot f''\right) e^{-iqr_j} e^{-\frac{\sigma_j^2 q^2}{2}} e^{-B_j q^2} \cdot C_R \tag{1}$$

where $j$ is the index of each atom, $q$ is the momentum transfer, $f_j$ is atomic form factor, $f'$, $f''$ is the energy dependent atomic form factor, and $B_j$ is the Debye Waller factor. Note that $\sigma_j$ is the standard deviation of each atomic layer distribution of and $C_R$ is the roughness factor [25–31] of the sample. Instead of controlling the roughness factor, the contribution of islands and domains with three buffer layers related to the graphene has been considered [See Section on the CTR analysis and Fig. S1 in the Supplementary Information].

Figure 3(b) shows the quantitative analysis results by fitting the model system. For the fitting analysis, we adjusted the occupancy and displacement of bulk Si and C atoms in the top-level SiC unit cell, $Si_{Bulk}$ and $C_{Bulk}$, respectively, compatible with the reference values [13]. The buffer layer was formed between the SiC substrate and graphene and is separated from each other [32,33]. The buffer layer was reconstructed and composed of two carbon atoms S1 and S2 located at 2.31 ± 0.3 Å and 1.91 ± 0.3 Å, respectively. The occupancy values of S1 and S2 were 1 and 0.89, respectively. The model had bilayer graphene in which the first graphene G1 and the second graphene G2 were positioned at 5.51 and 8.71 Å, respectively, as shown in Fig. 3(c). Our model had three domains consisting of buffer-only, monolayer graphene, and bilayer graphene with coverages of 0.23, 0.47, and 0.3, respectively. The G1 and G2 layers showed reasonable positions and distributions of 0.23 and 0.22 Å, respectively. The electron density profiles obtained from the fitting reveals the detailed occupancy and displacement of each element [see Fig. S1 in the Supplementary Information].

Figure 3(c) represents the out-of-plane lattice parameters of the bilayer $VSe_2$, graphene layer, and SiC substrate. Specifically, the center positions of the first and second $VSe_2$ layers were located at 4.625 Å and 10.640 Å from the last graphene layer, respectively. In the bilayer $VSe_2$ model, there are multiple $VSe_2$ domains with different thicknesses of $VSe_2$, mainly bilayer $VSe_2$ with coverage of 0.67 each. In the first $VSe_2$ layer, the occupancy of top and bottom Se layers labeled as Se(I) and Se(II), were 1 and 0.92, respectively. In addition, the occupancy of Se(I) and Se(II) in the second $VSe_2$ layer were 0.87 and 1, respectively. Surprisingly, we found that the top and bottom Se layer spacing of the first $VSe_2$ layer between layers was 2.87 Å. This is because the spacing between the two Se layers in the bulk sample is usually 3.10 Å, which is similar with the spacing between the top and bottom Se layers of the second $VSe_2$ layer. Compared with previously reported lattice parameters of 1T $VSe_2$, the out-of-plane lattice parameter of the first $VSe_2$ layer of the heterostructure was compressed to about 6.06 Å, while the lattice parameter of the second layer was nearly stretched by almost 6.13 Å. The average value of two layers was nearly identical with that of bulk 1T $VSe_2$. Theoretical and experimental studies have proposed that the distance between V and Se atom can lead to the emergence of ferromagnetic orders originated from the modification of electronic structure, orbital states, Se stoichiometry, and structure phase transitions.

To elucidate the in-plane alignment and bond length of the bilayer VSe$_2$ heterostructure, we performed phi-scan and off-specular CTR measurements, as shown in Fig. 4. However, the precise experimental observation of the in-plane alignment of vdW-based heterostructure is an important unresolved research topic. This is because the weak vdW interaction between two layers can lead to quasi-stable structures that differ significantly from the covalent bond-based heterostructures. Figure 4(a) shows a schematic diagram of the possible in-plane alignment of the VSe$_2$ layer and the SiC substrate in reciprocal space, showing approximately 30° rotation in the in-plane direction of VSe$_2$[100] relative to the SiC[10$\bar{1}$0] substrate. As shown in Fig. 4(b), we found that the peak spacing of the measured phi-scan at (0.518 0.518 2.4) reflection was about 60° in sixfold symmetry. Note that the average value of the full width half maximum (FWHM) values of reflections in the phi scan was 6.6°, which was in good agreement with the reported value (approximately 5°) measured by LEED [14].

A possible origin of the six-fold peaks from the heterostructure is the presence of domains in the graphene, 2H VSe$_2$ or 1T VSe$_2$ phase, with different in-plane orientations. To rule out the possibility of a graphene layer, we measured anomalous x-ray diffraction using Se K-edge. We clearly found absorption features near the Se K-edge, indicating the (0.518 0.518 2.4) reflection originated from the VSe$_2$ layer. As shown in Fig. 4(d), the calculated in-plane lattice parameter was 3.44Å, which was about 4.2% larger than that of bulk 1T VSe$_2$ phase. In-plane CTR simulations (red line in Fig. 4(d)) of bilayer VSe$_2$/bilayer graphene heterostructure grown on SiC substrate reproduce the experimental results obtained from the (0.518 0.518 2.4) reflection (black dots in Fig. 4(d)). The model included various in-plane orientations of 1T VSe$_2$ phase rather than the domains of 2H phase. Although the in-plane CTR measurements and analyses on the vdW heterostructures were so challengeable due to inherent weak and broad signals associated with the low dimension, CTR have provided new opportunities to identify atomic-level interfacial structures and to explore the intrinsic physical properties of vdW heterostructures.

**4. Conclusions**

In summary, we resolved the atomic-level structural properties of the bilayer $VSe_2$/bilayer graphene (BLG) heterostructures using crystal truncation rod (CTR) analysis. CTR analysis was able to address one-dimensional atomic arrangement of the $VSe_2$/BLG heterostructures. We confirmed the bilayer $VSe_2$ layer was a 1T phase with different in-plane domains. Furthermore, we quantitatively defined the electron density profiles leading to the atomic-level positions of elements and out-of-plane and in-plane lattice parameters of each layer of the heterostructure. CTR analysis provides a novel experimental method to directly measure detailed structural properties of van der Waals heterostructure systems and explore their intriguing functionalities.

**Acknowledgments**

The work was supported by the Chung-Ang University Graduate Research Scholarship in 2021 (awarded to R.K.) and was supported by Basic Science Research Programs through the National Research Foundation of Korea (Grants No. No. 2020R1C1C1012424). This work was supported by MOLIT as [Innovative Talent Education Program for Smart City].

**Figure Captions**

**Figure 1** | VSe$_2$ heterostructures. (a) Schematic of bilayer VSe$_2$/bilayer graphene (BLG) grown on 6H-SiC substrate. (b) Top and side view of 1T VSe$_2$ bulk structure and its lattice parameters. (c) X-ray diffraction (XRD) using a laboratory source of bilayer VSe$_2$/BLG heterostructure samples grown on 6H-SiC substrate (red line) and 6H-SiC substrate (black line). The index was assigned to the reciprocal lattice units (r. l. u.) of the 6H-SiC substrate.

**Figure 2** | Crystal truncation rod (CTR) scattering of VSe$_2$/BLG heterostructure. (a) CTR measurements of monolayer (blue) and bilayer (red) VSe$_2$ heterostructures measured along the out-of-plane direction, 00L. The (0 0 L) reflection of VSe$_2$ and the image of two-dimension (2D) area detector were indicated by inverted triangles and arrows, respectively. (b) 2D image of a monolayer VSe$_2$/BLG heterostructure sample measured at L=5.87 represents the CTR (red box) and Bragg peak of the substrate (white box). (c) 2D image measured at L=7.73 shows a representative anti-Bragg peak of our sample. (d) 2D image of monolayer VSe$_2$/BLG heterostructure sample measured at L=9.4.

**Figure 3** | CTR analysis of bilayer VSe$_2$/BLG heterostructure based on the model system. (a) CTR measurement data (black circle) and fitting result based on our model (red line). Green dashed line is a representative simulation based on lattice parameters of bulk 1T VSe$_2$. (b) Detailed estimated layer spacing of each atomic layer. (c) Out-of-plane lattice parameters of Buff., G1, G2, and VSe$_2$, which were the C buffer layer on the SiC substrate, first graphene, second graphene, VSe$_2$ layers, respectively. The blue and brown lines correspond to the Si and C layers, respectively. Error bars are the average distributions of domains.

**Figure 4** | Off-specular CTR measurements of bilayer VSe$_2$/BLG heterostructures for in-plane lattice

parameters. (a) A schematic of the reciprocal lattice vectors, a* and b*, of SiC and VSe$_2$ layers. (b) Phi scan measurement taken at (0.518, 0.518, 2.4). (c) Resonant energy scan measured at (0.518, 0.518, 2.4) near Se K edge. (d) Off-specular CTR measurement along (0.518 0.518 L) and a simulation result based on the model of bilayer VSe2/BLG heterostructures, denoted by black dot and red line, respectively. Gray dashed line is the noise level.

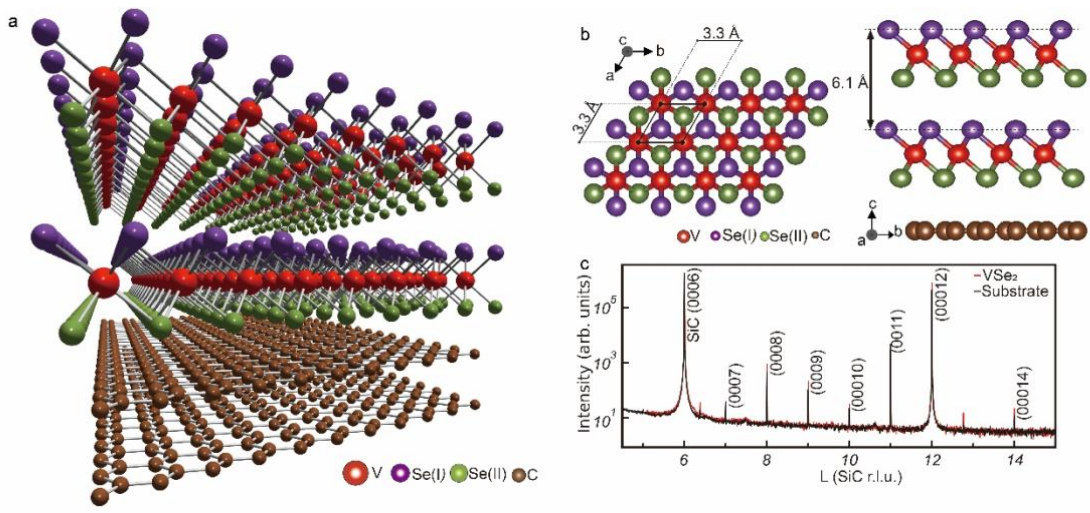

R. Kim *et al.*, Fig. 1.

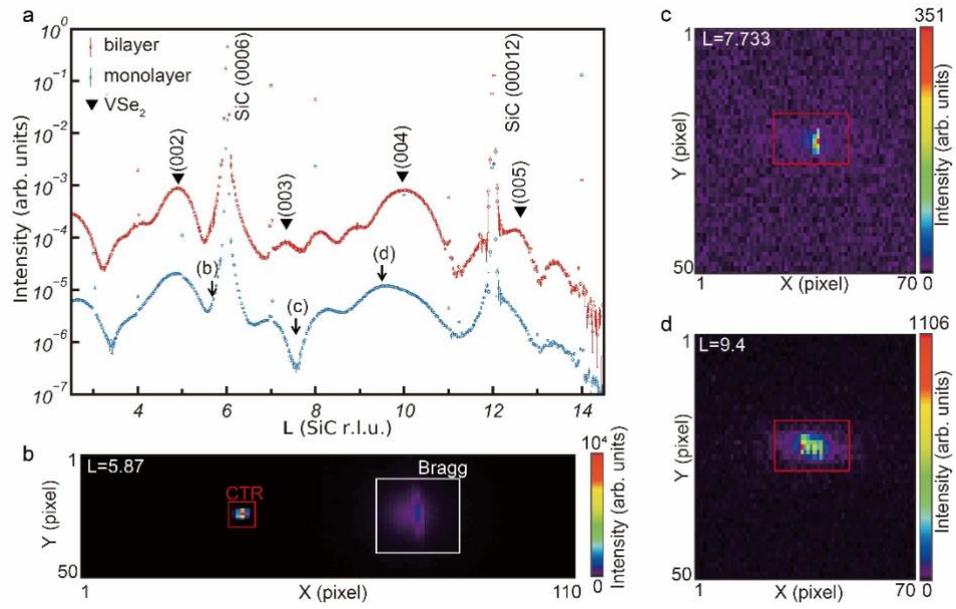

R. Kim *et al*., Fig. 2.

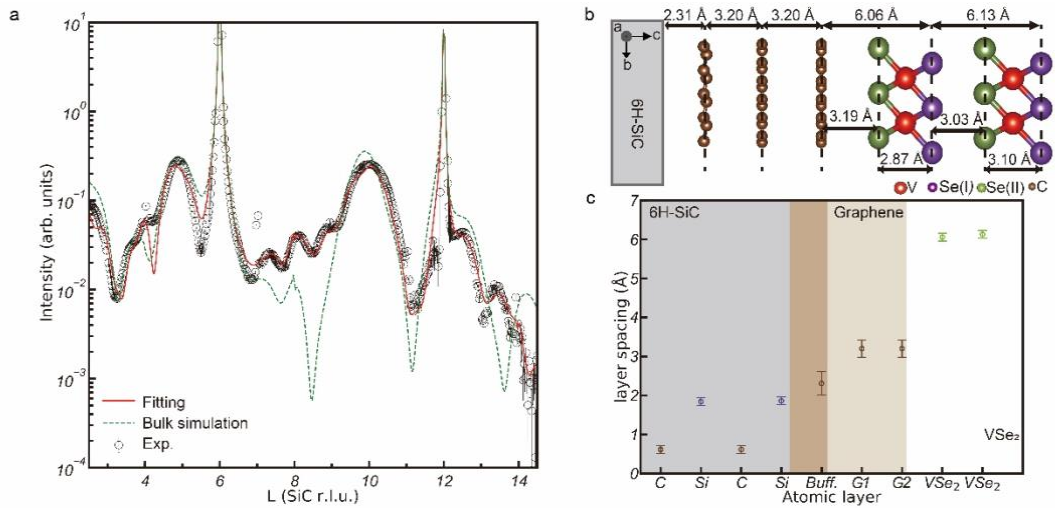

R. Kim *et al*., Fig. 3.

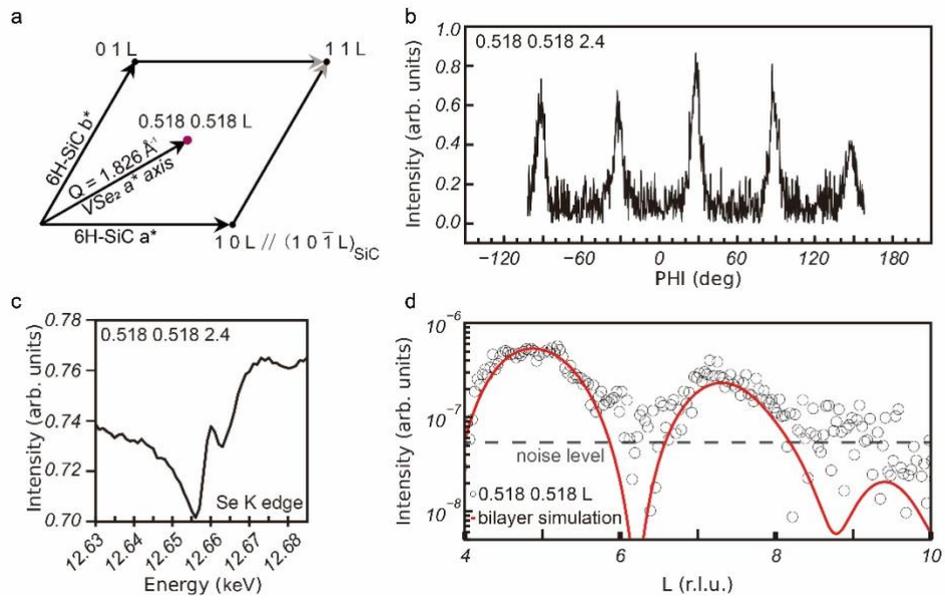

R. Kim *et al.*, Fig. 4.